\documentclass[aps,twocolumn,english,prx,showpacs,superscriptaddress,longbibliography]{revtex4-1}
\usepackage[colorlinks=true,urlcolor=blue,citecolor=blue,linkcolor=blue]{hyperref}

\usepackage[T1]{fontenc}
\usepackage[latin9]{inputenc}
\usepackage{amssymb}
\usepackage{graphicx}
\usepackage{amsmath,color}
\usepackage{mathrsfs}
\usepackage{float}
\usepackage{indentfirst}
\usepackage{subfigure}
\usepackage{multirow}
\usepackage{tabu}
\usepackage{booktabs}
\usepackage{txfonts}
\usepackage{amsmath,amssymb,bbm}

\usepackage[euler]{textgreek}

\usepackage{graphicx}
\usepackage{bm}

\usepackage[dvipsnames]{xcolor}
\usepackage{soul}

\newcommand{\mathd}{\mathrm{d}}
\newcommand{\mathi}{\mathrm{i}}

\begin{document}

\preprint{APS/123-QED}

\title{Quantized thermal Hall conductance from edge current calculations in lattice models}%

\author{Wei Tang}
\author{X.~C. Xie}
 \affiliation{International Center for Quantum Materials, School of Physics, Peking University, Beijing 100871, China}
\author{Lei Wang}
 \email{wanglei@iphy.ac.cn}
 \affiliation{Beijing National Lab for Condensed Matter Physics and Institute of Physics, Chinese Academy of Sciences, Beijing 100190, China}
 \affiliation{CAS Center for Excellence in Topological Quantum Computation, University of Chinese Academy of Sciences, Beijing 100190, China}
 \affiliation{Songshan Lake Materials Laboratory, Dongguan, Guangdong 523808, China} 
\author{Hong-Hao Tu}
 \email{hong-hao.tu@tu-dresden.de}
 \affiliation{Institute of Theoretical Physics, Technische Universit\"at Dresden, 01062 Dresden, Germany}

\date{\today}

\begin{abstract}
The quantized thermal Hall effect is an important probe for detecting chiral topological order and revealing the nature of chiral gapless edge states. 
The standard Kubo formula approach for the thermal Hall conductance $\kappa_{xy}$ based on the linear-response theory faces difficulties in practical application due to the lack of a reliable numerical method for calculating dynamical quantities in microscopic models at finite temperature. 
In this work, we propose an approach for calculating $\kappa_{xy}$ in two-dimensional lattice models displaying chiral topological order. 
Our approach targets at the edge current localized at the boundary which involves only thermal averages of local operators in equilibrium, thus drastically lowering the barrier for the calculation of $\kappa_{xy}$.
We use the chiral $p$-wave superconductor (with and without disorder) and the Hofstadter model as benchmark examples to illustrate several sources of finite-size effects, and we suggest the infinite (or sufficiently long) strip as the best geometry for carrying out numerical simulations.
\end{abstract}

\maketitle


\section{Introduction}

In recent years there has been a prodigious interest in non-Abelian anyons emerging in condensed matter systems, which are of great significance in the contexts of both fundamental science and applied subjects{~\cite{nayak2008non}}. 
These non-Abelian anyons, which appear as quasiparticles in systems with topological order, allow braiding manipulations---i.e., exchanging these quasiparticles drives the system into a distinct state---and upon these braiding processes lays the foundation for fault-tolerant quantum computation{~\cite{kitaev_fault-tolerant_2003,nayak2008non}}.
Originating from the fractional quantum Hall system with filling factor $\nu=5/2${~\cite{moore_nonabelions_1991, read2000paired}}, the search for non-Abelian anyons also proceeds to other quantum systems, such as superconductor-semiconductor heterostructures{~\cite{lutchyn2018majorana}} and chiral spin liquids{~\cite{kitaev_anyons_2006}}. 
Despite extensive theoretical and experimental effort, a completely confirmed experimental observation of non-Abelian anyons still remains elusive.

In realistic materials, the quantized thermal Hall effect could serve as a strong indicator of the existence of non-Abelian anyons. 
For chiral topological systems, there exist gapless edge modes at the boundary that, at the low-energy limit, are generally described by chiral conformal field theories (CFTs) in $(1+1)$ dimensions.
The energy current carried by the chiral edge mode has a universal form{~\cite{cappelli_thermal_2002}} $I(T) = (\pi k_B c_-/12 \hbar) T^2$, where $k_B$, $\hbar$, and $T$ denote the Boltzmann constant, the reduced Planck constant, and the temperature, respectively. 
Here $c_-$ is the chiral central charge, which characterizes the edge CFT and reflects the topological properties in the bulk, can be directly measured in a thermal Hall experiment, i.e., $\hbar \kappa_{x y}/k_B T = \pi c_-/6${~\cite{kane_quantized_1997,cappelli_thermal_2002}}, where $\kappa_{x y}$ is the thermal Hall conductance. 
Specifically, if the edge mode includes a Majorana branch, the Majorana edge mode would contribute a chiral central charge $c_-=1/2$.
In this regard, a half-integer quantized thermal Hall conductance provides strong evidence for the existence of the Majorana edge mode and non-Abelian anyon excitations in the bulk.
Very recently, the anticipated half-integer thermal Hall conductance is observed in the $\nu=5/2$ quantum Hall system{~\cite{banerjee_observation_2018}}, 
as well as the field-induced disordered state in the Kitaev material $\alpha$-RuCl$_3${~\cite{kasahara_majorana_2018}}. 

In spite of the remarkable experimental progress, methods for a direct calculation of the thermal Hall conductance from microscopic Hamiltonians are circumscribed, which poses an obstacle to the theoretical understanding of the experimental observations. 
For example, despite the observation of the half-integer quantized thermal Hall conductance in $\alpha$-RuCl$_3$, the corresponding theoretical interpretation is still under debate. 
The microscopic model describing the magnetic properties of $\alpha$-RuCl$_3$ remains obscure (see, e.g., \cite{trebst_kitaev_2017,janssen2019heisenberg}), and it is largely unclear whether the experimentally observed field-induced disordered state is adiabatically connected to the non-Abelian spin liquid with dominant Kitaev interaction perturbed by a magnetic field.
One of the important reasons for these obscurities is attributed to the lack of a suitable theoretical toolbox for handling these problems---the existing numerical calculations of the thermal Hall conductance rely on the usage of the Kubo formula{~\cite{kubo_statistical-mechanical_1957,luttinger_theory_1964}}, which requires an evaluation of dynamical quantities, which are commonly unattainable in interacting systems.
These calculations are limited, therefore, to the cases in which a mapping of the system to a quadratic fermion model is possible{~\cite{nasu_thermal_2017,metavitsiadis_thermal_2017}}, or to small-sized systems that are within the capability of exact diagonalization. 
Moreover, the numerical application of the Kubo formula in thermal Hall problems is plagued by the subtle magnetization correction that originates from circulating currents in the system{~\cite{cooper_thermoelectric_1997,qin_energy_2011,nomura_cross-correlated_2012,stone_gravitational_2012,sumiyoshi_quantum_2013,vinkler_bulk_2019}}, and the existing analytic solution{~\cite{qin_energy_2011,tatara_thermal_2015}} to this correction is rather difficult to use in numerics, especially in the presence of interactions.
For these reasons, a direct calculation of the thermal Hall conductance from interacting systems remains elusive.

In this work, we propose a method to directly address the edge modes of chiral topological systems and compute the quantized thermal Hall conductance from the edge currents, as motivated by Kitaev's work {\cite{kitaev_anyons_2006}}. 
In contrast to previous work using the Kubo formula, our edge current approach only involves thermodynamic calculations of the locally defined operators, which are of much less computational cost. 
Furthermore, this approach is rather straightforward and thus is free from impediments such as energy magnetization corrections.
In these respects, it has the potential to be applied to interacting systems with prevailing numerical algorithms.

As a preliminary study, we investigate the performance of the edge current approach using several non-interacting fermion systems as benchmark examples.
By analyzing the finite-size effects in this approach, including the discretization of the edge spectrum and the overlap between edge modes, we suggest the finite-width long strip as an optimal choice of the lattice geometry when applying it to interacting systems, which might be within the scope of the state-of-the-art tensor-network algorithms.
The edge current approach using this suggested lattice geometry is further tested in the Hofstadter model and the disordered $p$-wave superconductor.

This paper is organized as follows. 
In Sec.~\ref{sec:method_sec}, we present our edge current approach to thermal Hall conductance, and we describe the applicability of this approach.
In Sec.~\ref{sec:discussions}, we show preliminary numerical results in the chiral $p$-wave superconductor, discuss two major finite-size effects in this approach, and suggest an optimal lattice geometry for this approach. 
In Sec.~\ref{sec:applications}, we apply our method to the Hofstadter model and the disordered chiral $p$-wave superconductor. 
Finally, in Sec.~\ref{sec:conclusion_and_outlook} we summarize our work and give an outlook. 
In Appendix \ref{app:finite-size-correction-oneoverL2}, we review the renowned universal $T^2$-dependence of the edge energy current and also derive the analytical form of the ground-state contribution to the energy current.
In Appendix \ref{appendix:majorana-basis}, we provide details on how our approach is performed in free-fermionic systems.

\section{Method}\label{sec:method_sec}

\begin{figure}[!htb]
  \resizebox{8.5cm}{!}{\includegraphics{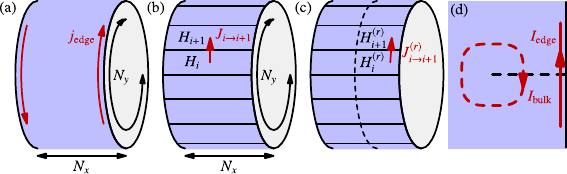}}
  \caption{(a) A schematic plot showing the edge thermal current flowing at the boundaries of a chiral topological system.
  (b) The energy current operator along one direction in a two-dimensional lattice. 
  (c) The right part of the spatially divided energy current operator. 
  (d) A schematic figure showing the edge current and bulk-state contributions contained in $\langle \hat{j}^{(r)} \rangle_T$.
  \label{fig:construction_curr_operator}}
\end{figure}

We begin by introducing our edge current approach for the thermal Hall conductance.
We put the system upon which we focused in our work---the gapped chiral topological system---on a cylinder, in order to introduce system boundaries and get access to the gapless edge modes.
When the temperature of the system is much smaller than the bulk gap $\Delta_{\mathrm{bulk}}$, the energy current carried by the chiral edge mode, from which the quantized thermal Hall effect originates
\footnote{In $U(1)$-conserved systems, the heat current $J_Q$ and the energy current $J_E$ are conceptually different, and they are related to each other by $J_Q=J_E- \mu J_N$, where $J_N$ and $\mu$ represent the $U(1)$ current and the chemical potential of the system, respectively.
In these systems, however, it is always legitimate to shift the chemical potential to zero by absorbing $-\mu \hat{N}$ into the Hamiltonian $\hat{H}$, i.e., $\hat{H} \rightarrow \hat{H} - \mu \hat{N}$, where $\hat{N}$ is the particle number operator. 
Along these lines, we hereby identify the energy current and the thermal current in this paper.},
can naturally be used to compute the thermal Hall conductance [see Fig.~\ref{fig:construction_curr_operator}(a)]
\begin{equation}
    \kappa_{x y} = \frac{\mathd j_{\mathrm{edge}}(T)}{\mathd T} \text{ ~for } T \ll \Delta_{\mathrm{bulk}}.
    \label{eq:starting_point_of_method}
\end{equation}
Here $j_{\mathrm{edge}}(T)$ denotes the edge current density in the thermal equilibrium state at the temperature $T$, which is predicted to be of the universal form{~\cite{cappelli_thermal_2002}}
\begin{equation}
    j_{\mathrm{edge}}(T) = \frac{\pi c_-}{12} T^2 \text{ ~for } T \gg v/N_y, \label{eq:cft_IedgeT}
\end{equation}
where we have set $k_B=\hbar=1$.
Here $c_-$, $v$, and $N_y$ denote the chiral central charge, the velocity of the edge mode and the circumference of the cylinder, respectively. 
A straightforward combination of Eqs.{~\eqref{eq:starting_point_of_method}} and {\eqref{eq:cft_IedgeT}} gives the quantized thermal Hall conductance
\begin{equation}
    \frac{\kappa_{x y}}{T} = \frac{\pi c_-}{6},
    \label{eq:quantized_thermal_hall_conductance_expr}
\end{equation}
where $T$ satisfies $T \ll \Delta_{\mathrm{bulk}}$ and $T \gg v / N_y $.

Despite the simplicity of this scheme, it is far from obvious how to obtain $j_{\mathrm{edge}}(T)$ in numerical calculations.
For this purpose, we first introduce the definition of local energy current in the lattice system.
For a lattice model constituted by local Hamiltonian terms $\hat{H} = \sum_m \hat{H}_m$, according to the Heisenberg equation
\begin{equation}
    \frac{\mathd \hat{H}_n}{\mathd t} = - \mathi [\hat{H}_n, \hat{H}] = -\sum_m \mathi [\hat{H}_n, \hat{H}_m],
\end{equation}
one may define the local current flowing from $\hat{H}_m$ to $\hat{H}_n$ as 
\begin{equation}
    \hat{J}_{m\rightarrow n} = \mathi [\hat{H}_m, \hat{H}_n]. \label{eq:local-current-def-firstversion}
\end{equation} 
In practice, however, there may exist multiple ways to define such local energy currents in a lattice system. 
First, the choice of partitioning the total Hamiltonian into local Hamiltonian terms is not unique. 
Moreover, given a choice of local Hamiltonian terms, Eq.{~\eqref{eq:local-current-def-firstversion}} may be only one of the possible definitions---in two dimensions, there may exist more than one definitions of the local currents that satisfy the energy conservation\cite{qin_energy_2011,metavitsiadis_thermal_2017}.
These ambiguities in the definition of the local energy current may call into question the reliability of calculations based on this definition, especially when the Hamiltonian is rather complicated---for example, there may exist multiple-site interactions in the system. 

Nevertheless, given a physically correct approach, the calculation results of the thermal Hall conductance---a measurable physical quantity---should not depend on the choice of conventions. 
Indeed, in the following, we introduce a suitable setup, under which the effect of the convention differences is negligible.

To start with, we view the two-dimensional system as a quasi-one-dimensional system by slicing the two-dimensional lattice into layers, as schematically depicted in Fig.{~\ref{fig:construction_curr_operator}(b)}.
We presume that only local Hamiltonians at the nearest-neighboring layers have non-vanishing commutators, i.e., $[\hat{H}_m, \hat{H}_n] \neq 0$ only when $|m-n|=1$, where $\hat{H}_{m}$ denotes the local Hamiltonian at the $m$-th layer.
This condition can be achieved with a proper choice of Hamiltonian term at each layer.
In this simple one-dimensional system, the energy current density along the circumferential direction is apparently given by 
\begin{equation}
  \hat{j} = \frac{1}{N_y} \sum_n \hat{J}_{n \rightarrow n + 1} =
  \frac{\mathrm{i}}{N_y} \sum_n [\hat{H}_n, \hat{H}_{n + 1}],
  \label{eq:quasi-1d-current-op}
\end{equation}
where we have set the length of each layer to be one unit length in the circumferential direction [see Fig.{~\ref{fig:construction_curr_operator}(b)}].

Next, in order to investigate the edge currents of the system, we spatially divide the layers into two parts, $\hat{H}_n = \hat{H}_n^{(l)} + \hat{H}_n^{(r)}$, where $\hat{H}_n^{(l / r)}$ is on the left/right side [see Fig.~\ref{fig:construction_curr_operator}(c)]. 
The current operator $\hat{J}_{n\rightarrow n + 1}$ becomes
\begin{eqnarray}
  \hat{J}_{n \rightarrow n + 1} & = & \mathrm{i} [\hat{H}_n^{(l)},
  \hat{H}^{(l)}_{n + 1}] + \mathrm{i} [\hat{H}_n^{(l)}, \hat{H}^{(r)}_{n + 1}]
  \nonumber\\
  &  & + \mathrm{i} [\hat{H}_n^{(r)}, \hat{H}^{(l)}_{n + 1}] + \mathrm{i}
  [\hat{H}_n^{(r)}, \hat{H}^{(r)}_{n + 1}] . 
\end{eqnarray}
Since the edge modes are exponentially localized at the boundaries, it is legitimate to concentrate on $\hat{J}_{n \rightarrow n + 1}^{(r)} \equiv \mathrm{i} [\hat{H}_n^{(r)}, \hat{H}^{(r)}_{n + 1}]$ which covers the edge current contributions at the right boundary [see Fig.~\ref{fig:construction_curr_operator}(c)]. 
In this regard, we introduce the energy current density operator in the right part
\begin{equation}
  \hat{j}^{(r)} = \frac{1}{N_y} \sum_n \hat{J}^{(r)}_{n \rightarrow n + 1} =
  \frac{\mathrm{i}}{N_y} \sum_i [\hat{H}_n^{(r)}, \hat{H}^{(r)}_{n + 1}],
  \label{eq:j-r-def}
\end{equation}
whose thermal average $\langle\hat{j}^{(r)}\rangle_T$ contains all the edge current contributions at the right boundary.
Here $\langle\hat{j}^{(r)}\rangle_T$ is given by $\langle\hat{j}^{(r)}\rangle_T = (1/Z)\mathrm{Tr}(\mathrm{e}^{-\beta \hat{H}} \hat{j}^{(r)})$, where $Z$ represents the partition function, and $\beta = 1 / T$. 
It is apparent that $\langle \hat{j}^{(r)} \rangle_T$ is \emph{not} solely contributed by the edge current---it also contains contributions from bulk states. 
These bulk-state contributions may come from the circulating bulk currents [see Fig.~\ref{fig:construction_curr_operator}(d)]. 
Furthermore, due to the spatial separation $\hat{H}_n = \hat{H}_n^{(l)} + \hat{H}_n^{(r)}$, $\langle \hat{j}^{(r)} \rangle_T$ may also contain bulk-state contributions at the dividing point, and for different ways of dividing the local Hamiltonians in the central region, these bulk contributions are different.
Nevertheless, if we restrict the temperature to be much smaller than the bulk gap (as in Eq.~\eqref{eq:starting_point_of_method}), the bulk states with energies higher than the bulk gap are suppressed---in this case, the bulk-state contributions contained in $\langle \hat{j}^{(r)} \rangle_T$ mostly come from the ground state and are hence nearly independent of the temperature. 
Along these lines, and combining Eq.~\eqref{eq:starting_point_of_method}, the thermal Hall conductance can be calculated as 
\begin{equation}
  \kappa_{x y} = \frac{\mathd j_{\mathrm{edge}}}{\mathd T} = \frac{\mathd
  \langle \hat{j}^{(r)} \rangle_T}{\mathd T}, \label{eq:dIdT-to-kappa}
\end{equation}
where we have restricted the temperature to $T \ll \Delta_{\mathrm{bulk}}$.

Equation (\ref{eq:dIdT-to-kappa}) is the central result of our work. 
In numerical applications, one needs to calculate $\langle \hat{j}^{(r)} \rangle_T$ for several temperatures and make use of the numerical differentiation.
An alternative recipe is to subtract the ground-state contribution in $\langle \hat{j}^{(r)} \rangle_T$, i.e., to make use of the $T^2$-dependence of the edge current and calculate $\kappa_{x y}$ as 
\begin{eqnarray}
  \kappa_{x y} & = & \frac{2}{T} \left( j_{\mathrm{edge}} (T) -
  j_{\mathrm{edge}} (T = 0) \right) \nonumber\\
  & = & \frac{2}{T} (\langle \hat{j}^{(r)} \rangle_T - \langle \hat{j}^{(r)}
  \rangle_{T = 0}), \label{eq:gs_subtracted_jr}
\end{eqnarray}
where $T \ll \Delta_{\mathrm{bulk}}$, and bulk-state contributions in $\langle \hat{j}^{(r)} \rangle_T$ and $\langle \hat{j}^{(r)} \rangle_{T = 0}$ cancel each other out.
However, we need to point out that Eq.{~\eqref{eq:gs_subtracted_jr}} suffers from a finite-size correction that comes from the subtracted ground-state edge current $j_{\mathrm{edge}} (T = 0) \sim O(1/N_y^2)$. 
The analytical form of this finite-size correction is derived in Appendix \ref{app:finite-size-correction-oneoverL2}.

We emphasize that Eq.~(\ref{eq:dIdT-to-kappa}) only involves the thermal averages over locally defined quantities, and it requires much less numerical cost compared to the conventional approach based on the Kubo formula{~\cite{luttinger_theory_1964,qin_energy_2011,nomura_cross-correlated_2012,sumiyoshi_quantum_2013}}, which relies on the evaluation of the dynamics of the system. 
Furthermore, our approach is free from the subtle issue of energy magnetization correction{~\cite{qin_energy_2011}}. 
As a trade-off, this approach is restricted to gapped chiral topological systems, and the temperature must be much lower than the bulk gap.
In this regard, the edge current approach cannot be applied to the thermal Hall effect contributed by magnons{~\cite{katsura_theory_2010,onose_observation_2010,matsumoto_theoretical_2011,matsumoto_rotational_2011}} or phonons{~\cite{strohm_phenomenological_2005,sheng_theory_2006}}, where the bulk states are gapless and the corresponding thermal Hall conductance is generally nonquantized.

\section{Discussions}\label{sec:discussions}

Despite the simplicity of Eq.~(\ref{eq:dIdT-to-kappa}), this equation gives the anticipated quantized thermal Hall conductance only when the following conditions are satisfied.
First, the temperature $T$ must satisfy $T \ll \Delta_{\mathrm{bulk}}$ and $T \gg v / N_y$.
In addition, the length of the cylinder, $N_x$, should be large enough so that the overlap between edge modes is minimized.
In these respects, there are restrictions on the choice of the system size.
For non-interacting fermion systems, one can calculate Eq.~\eqref{eq:dIdT-to-kappa} to reasonably large system sizes (see Appendix \ref{appendix:majorana-basis} for details of the calculation).
Hence, in this section, using the $p+\mathrm{i} p$ superconductor on the square lattice as a platform, we numerically investigate the finite-size effects arising in the edge current approach and look for an optimal lattice geometry for this calculation.
The Hamiltonian of the $p+\mathrm{i} p$ model is given by
\begin{eqnarray}
  H & = & \sum_{m, n} \left\{ - t \left( f_{m + 1, n}^{\dagger} f_{m, n} +
  f_{m, n + 1}^{\dagger} f_{m, n} + \mathrm{h.c.} \right) - \mu f_{m,
  n}^{\dagger} f_{m, n} \right. \nonumber\\
  &  & + \left( \Delta f_{m + 1, n}^{\dagger} f_{m, n}^{\dagger} + \mathrm{i}
  \Delta f_{m, n + 1}^{\dagger} f_{m, n}^{\dagger} + \mathrm{h.c.} \right\}, 
  \label{eq:hamilt-ppip}
\end{eqnarray}
where $f_{m,n}$ is the fermion operator, with $m,n$ representing the site index.
$t$, $\mu$, and $\Delta$ correspondingly represent the hopping parameter, the chemical potential, and the pairing potential.
We set $t > 0$ and $\Delta > 0$.
When $-4 t < \mu < 0$ and $0 < \mu < 4 t$, the $p + \mathrm{i} p$ model is in the gapped topological phase (also known as the ``weak pairing'' phase), where the sign of $\mu$ determines the chirality,
while for $\mu < -4 t$ or $\mu > 4 t$, the system is in the trivial phase (also known as the ``strong pairing'' phase)~\cite{read2000paired}.
When the system is in the topological phase, a chiral Majorana edge mode exists at the boundary and leads to a chiral central charge $c_-= \pm 1/2$, where the sign is determined by the chirality of the edge mode.
In the following, we concentrate on the topological phase corresponding to $0 < \mu < 4t$, and we demonstrate the performance of the edge current approach in the case of various system sizes. 

\subsection{Restrictions on $N_y$: discretization of the edge spectrum and bulk contributions}

Let us first explore the requirements on the circumference of the system. 
As previously mentioned, Eq.~\eqref{eq:dIdT-to-kappa} gives the anticipated quantized thermal Hall conductance only when the conditions $T \gg v / N_y$ and $T \ll \Delta_{\mathrm{bulk}}$ are satisfied.
The first condition $T \gg v / N_y$ arises due to the discrete edge spectrum---the temperature should be much larger than the edge spectrum discretization $v/N_y$; otherwise, the edge spectrum appears as a few separate energy levels and the prediction from the continuous field theory will break down.
The second condition $T \ll \Delta_{\mathrm{bulk}}$ is imposed to prevent bulk-state contributions.
Apparently, to ensure the existence of a temperature regime that satisfies the two conditions simultaneously, we should choose a sufficiently large circumference $N_y$.

In the $p+\mathrm{i}p$ superconductor, whose Hamiltonian is given by Eq.~\eqref{eq:hamilt-ppip}, we numerically investigate the constraint on the circumference of the system. 
In Eq.~\eqref{eq:hamilt-ppip}, we fix the hopping parameter and the pairing potential as $t=\Delta=1.0$.
We set the length of the system to be $N_x=32$, which is large enough to avoid the overlap between edge modes at opposite boundaries.
The circumference of the system varies between $N_y=20$ and $N_y=200$, which leads to different edge spectrum discretizations.
Meanwhile, we adjust the bulk gap by changing the chemical potential between $\mu=1.0$ and $\mu=0.5$. 

\begin{figure}[!htb]
  \resizebox{8.6cm}{!}{\includegraphics{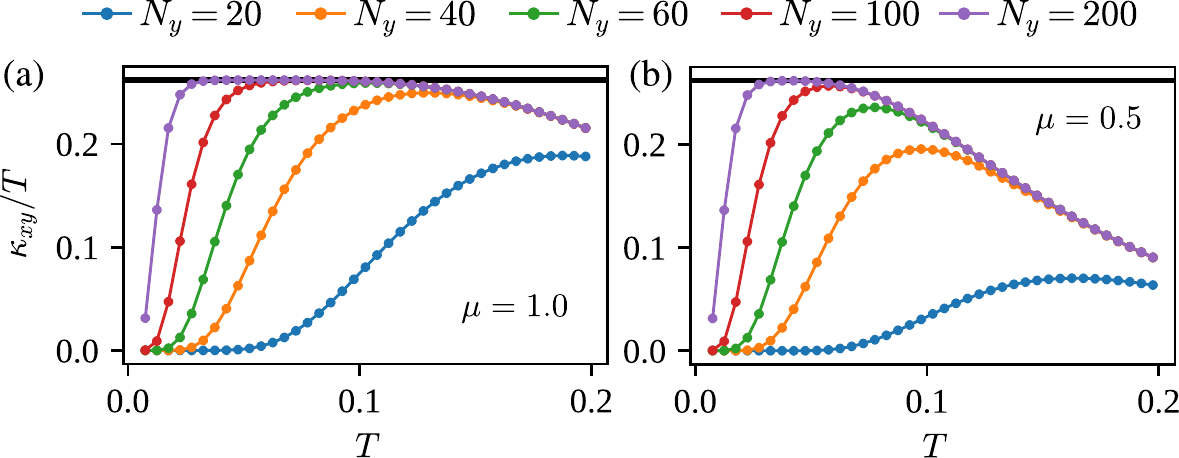}}
  \caption{$\kappa_{x y} / T$ calculated from the edge current in the $p + \mathrm{i} p$ superconductor with $t = \Delta = 1.0$ and $N_x = 32$. The circumference of the system $N_y$ varies from $N_y = 20$ to $N_y = 200$. 
  (a) and (b) shows the result for $\mu = 1.0$ ($\Delta_{\mathrm{bulk}}=1.0$) and $\mu = 0.5$ ($\Delta_{\mathrm{bulk}}=0.5$), respectively. 
  The expected quantization value $\pi c_- / 6 = \pi / 12$ is denoted by black horizontal lines.
 \label{fig:kappa-ppip-32xV-1x1}}
\end{figure}

Fig.~\ref{fig:kappa-ppip-32xV-1x1} shows the numerical results of the calculated $\kappa_{x y} / T$ from the edge current approach.
As can be observed in Fig.~\ref{fig:kappa-ppip-32xV-1x1}, if the system circumference is sufficiently large, the anticipated plateau appears.
Apart from the anticipated plateau, deviations from the quantization appear in both the low-temperature and high-temperature regimes.
By comparing results calculated with different system circumferences, one can verify that the low-temperature deviations mainly originate from the edge spectrum discretizations, since these deviations are reduced in the cases of larger circumferences.
On the other hand, by comparing Fig.~\ref{fig:kappa-ppip-32xV-1x1}(a) and (b), we find that for the case with a smaller bulk gap, the plateau shrinks evidently in the high-temperature regime, which is in agreement with our analysis---the edge current approach is affected by bulk-state contributions at high temperatures. 

The numerical results in Fig.~\ref{fig:kappa-ppip-32xV-1x1} also suggest that the required circumference of the system is very large, which is of the scale $N_y \sim 10^2$ (or probably even higher, if the bulk gap is smaller).

\begin{figure}[!htb]
  \resizebox{6cm}{!}{\includegraphics{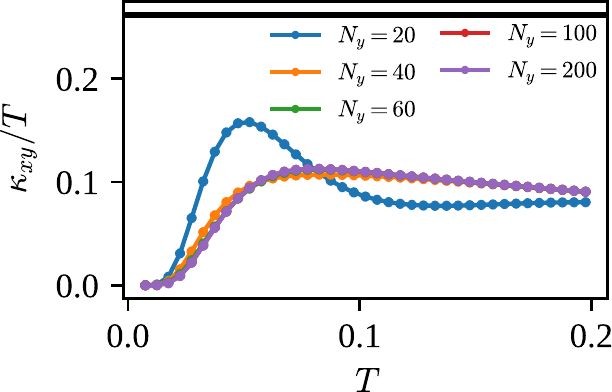}}
  \caption{$\kappa_{x y} / T$ calculated from edge current in the $p + \mathrm{i} p$ superconductor in the trivial phase. The parameters of the model are chosen as $t = \Delta = 1.0$ and $\mu = 4.1$. The system length is fixed as $N_x = 32$ while the circumference varies among $N_y = 20, 40, 60, 100, 200$. 
  The black horizontal line marks the quantization value $\pi / 12$ corresponding to $c_- = 1 / 2$. \label{fig:kappa-ppip-trivial}}
\end{figure}

As a comparison, we show the numerical result of $\kappa_{x y} / T$ in the gapped trivial phase in Fig.~\ref{fig:kappa-ppip-trivial}, where $t=\Delta=1.0$, $\mu=4.1$, and the bulk gap $\Delta_{\mathrm{gap}}=0.1$. 
In Fig.~\ref{fig:kappa-ppip-trivial}, the $\kappa_{x y}$ result vanishes at low temperatures and becomes finite when the temperature is high enough. 
This observation is consistent with our previous analysis: Due to the existence of the bulk gap, when $T \ll \Delta_{\mathrm{bulk}}$, the bulk contributions to $\langle \hat{j}^{(r)} \rangle_T$ are nearly independent of $T$, and the result becomes zero due to the absence of edge modes.
When the temperature is high, the bulk states are populated, which give rise to nonzero contributions to the numerical results of $\kappa_{x y}/T$.
Unlike the edge current, these bulk-state contributions show no apparent dependence on $N_y$ for large enough $N_y$, which may be used as a criterion to distinguish the edge and bulk currents.
This is in agreement with the results in Fig.~\ref{fig:kappa-ppip-32xV-1x1}, where $\kappa_{x y} / T$ in the high-temperature regime also becomes independent of $N_y$ for $N_y \geq 40$, since $\kappa_{x y} / T$ results at high temperatures are mainly contributed by bulk states.
The $\kappa_{x y} / T$ data in the trivial phase also shows a specious plateau, which mostly lies at relatively high temperature and corresponds to a nonquantized value. 
This suggests that essential prior knowledge of the chiral central charge or energy scale of the system is necessary; otherwise, this specious plateau in the trivial phase may lead to fallacious conclusions.

\subsection{Restrictions on $N_x$: overlap effect}

The other restriction on the geometry of the lattice comes from the overlap between edge modes. When deriving Eq.~(\ref{eq:dIdT-to-kappa}), it is assumed that the edge modes are well localized at the boundaries. In practice, however, the edge modes usually have finite decaying length $l_e$. By dimensional analysis, one can infer that the decaying length $l_e$ scales as $l_e \propto v / \Delta_{\mathrm{bulk}}$, where $v$ is the velocity of the edge mode.

When the system is deep in the topological phase, the bulk gap is usually rather large and the edge modes are well localized. As the system approaches the phase transition point, the $\Delta_{\mathrm{bulk}}$ decreases and the edge modes delocalize into the bulk. In this case, if the system length $N_x$ is not sufficiently large, the edge modes on opposite boundaries would have a large overlap with each other, which breaks the validity of Eq.~(\ref{eq:dIdT-to-kappa}).

\begin{figure}[!htb]
  \resizebox{8.6cm}{!}{\includegraphics{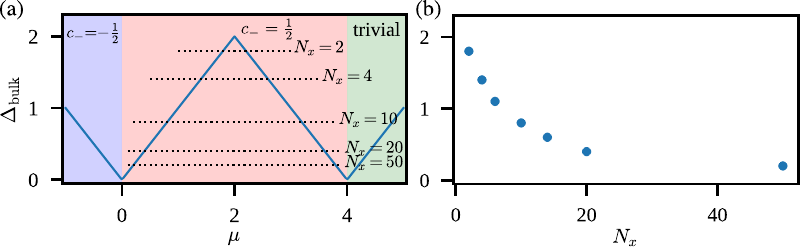}}
  \caption{The capability of the edge current approach for different choices   of $N_x$'s, represented by the minimal $\Delta_{\mathrm{bulk}}$ for each $N_x$ to obtain a satisfactory result of $c_-$. The minimal $\Delta_{\mathrm{bulk}}$ is (a) marked in the phase diagram of the $p + \mathrm{i} p$ model and (b) plotted with respect to $N_x$.
 \label{fig:Delta-bulk-vs-Nx}}
\end{figure}

Using the $p + \mathrm{i} p$ superconductor given by Eq.~(\ref{eq:hamilt-ppip}), we numerically investigate the overlap effect. By setting $t = \Delta = 1.0$, the velocity of the edge mode is fixed as $v = 2.0$, and the $\Delta_{\mathrm{bulk}}$ varies linearly with $\mu$ [see Fig.~\ref{fig:Delta-bulk-vs-Nx}(a)]. 
We adjust the decaying length of the edge mode by varying the chemical potential and thus the bulk gap $\Delta_{\mathrm{bulk}}$, and we test the performance of the edge current approach for different choices of $N_x$'s. 
For each $N_x$, the capability of the edge current approach is determined by a search for the minimal $\Delta_{\mathrm{bulk}}$ where the quantized plateau in $\kappa_{x y} / T$ can appear. The results are summarized in Fig.~\ref{fig:Delta-bulk-vs-Nx}.

From Fig.~\ref{fig:Delta-bulk-vs-Nx}, one can see that the required $N_x$ diverges as the bulk gap approaches zero, and when $N_x \leqslant 10$, the edge current approach can only capture the central region of the topological phase. 
This indicates that the edge current approach is capable of detecting the existence of a chiral topological phase, but it lacks the power to precisely determine topological phase transition points, where the bulk gap closes.

In the following, we concentrate on the cases when the overlap effect is not negligible but is of weak strength. We show that in this case, although the condition for Eq.~(\ref{eq:dIdT-to-kappa}) is violated, the edge current approach can still approximately give the quantized $\kappa_{x y} / T$.

The overlap between the edge mode induces a small gap $\Delta_{\mathrm{edge}}$ in the edge spectrum and drive the low-energy part of the edge spectrum apart from linearity. 
In this case, the low-temperature behavior of the numerical result for $\kappa_{x y} / T$ depends on the choice of system circumference $N_y$.
For systems with finite circumference $N_y$, the edge spectrum is discrete with size $\sim v/ N_y$.
For small $N_y$ which satisfies $\Delta_{\mathrm{edge}} \ll v / N_y$, when the temperature is in the vicinity of $T \sim \Delta_{\mathrm{edge}}$, the higher-energy excitations (with energies satisfying $E \geq \Delta_{\mathrm{edge}} + v/N_y \gg T$) are far from populated and only the lowest-energy excitations at $\Delta_{\mathrm{edge}}$ contribute to the thermal Hall conductance. 
As a result, a peak in $\kappa_{x y} / T$ would arise in the vicinity of $T \sim \Delta_{\mathrm{edge}}$. 
This large low-temperature peak can be remedied by choosing a larger $N_y$, and when the condition $\Delta_{\mathrm{edge}} \ll v / N_y$ is no longer satisfied, the low-temperature peak would gradually disappear. 
However, due to the non-linearity of the low-energy part of the spectrum, there will still be a weak deviation from the quantization value in this temperature regime. 
Despite these finite-size effects, one may expect that the edge spectrum remains linear at relatively higher temperatures and gives rise to an approximately quantized $\kappa_{x y} / T$.

We numerically verify the preceding analysis using the $p + \mathrm{i} p$ model with weak overlap effects. The overlap effect is gradually enhanced by adjusting the chemical potential $\mu$ towards the phase transition point. We sequentially calculate $\kappa_{x y} / T$ for $\mu = 0.9, 0.8, 0.7$. The length of the system is fixed as $N_x = 10$, while the system circumference varies among $N_y = 200, 400, 600$.

\begin{figure}[!htb]
  \resizebox{8.6cm}{!}{\includegraphics{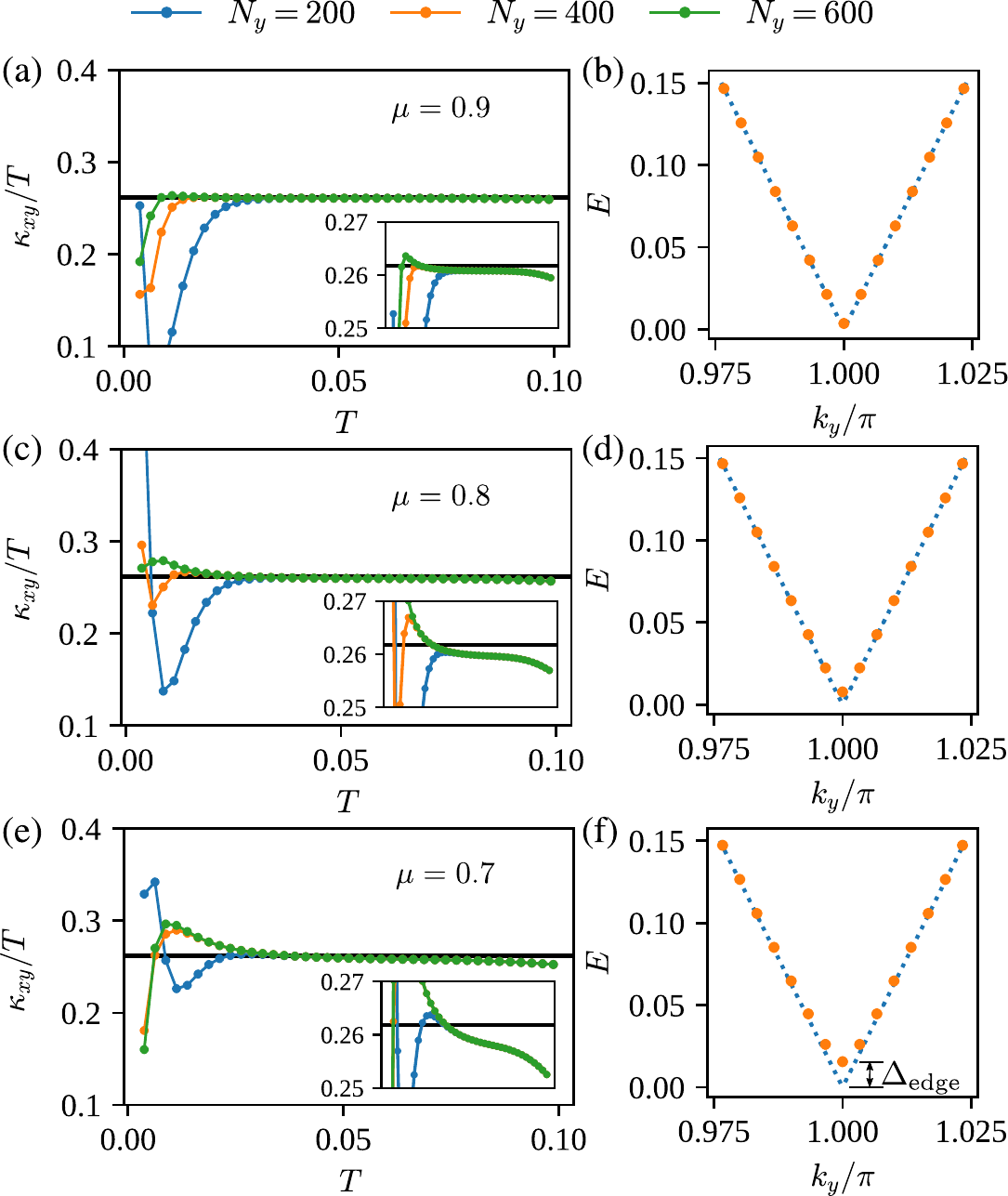}}
  \caption{$\kappa_{x y} / T$ calculated in the $p + \mathrm{i} p$ model using the edge current approach in the presence of the weak overlap effect. 
  We set $t = \Delta = 1.0$, $N_x = 10$ and $N_y = 200, 400, 600$. The numerical results of $\kappa_{x y} / T$ for $\mu = 0.9, 0.8, 0.7$ are shown, respectively, in (a)(c)(e), compared to the theoretically predicted quantized value $\pi / 12$ marked as black horizontal lines. 
  The insets in (a)(c)(e) zoom in on the data that are close to the quantization value. 
  As a comparison, the low energy part of the edge spectrum for $\mu = 0.9, 0.8, 0.7$ are respectively plotted in (b)(d)(f). 
  The corresponding energy spectrum and numerical results of $\kappa_{x y} / T$ share the same parameters, except that the edge spectrum is calculated only with $N_y = 600$.
  To make the edge gaps more clear, we mark perfect linear spectra as dashed lines in (b)(d)(f). 
  We mark the edge gap $\Delta_{\mathrm{gap}}$ explicitly in (f), while in (b)(d) the edge gaps also exist but are less visible.
  \label{fig:ppip-edge-overlap-invest}}
\end{figure}

Fig.~\ref{fig:ppip-edge-overlap-invest} shows the numerical results and the corresponding edge spectra. 
As indicated by the numerical results of $\kappa_{x y} / T$ in Figs.~\ref{fig:ppip-edge-overlap-invest}(a)(c)(e), the overlap effect arises for each choice of $\mu$ but has different strengths in different cases. 
This is in agreement with the edge spectra shown in Figs.~\ref{fig:ppip-edge-overlap-invest}(b)(d)(f), where the overlap effects---the small edge gap and the low-energy deviation from linearity---are gradually enhanced as the chemical potential approaches the phase transition point at $\mu = 0$. 
What is more, for relatively small $N_y = 200$, we observe the low-temperature peak in the $\kappa_{x y} / T$ result, which is attributed to the singular contribution of the lowest excitation state that arises in the temperature regime $T \sim \Delta_{\mathrm{edge}}$, as the condition $\Delta_{\mathrm{edge}} \ll v / N_y$ is satisfied. 
This finite-size artifact is remedied when we increase the system circumference to $N_y = 600$, and we are left with a small hump in $\kappa_{x y} / T$ originating from the low-energy nonlinearity of the edge spectrum. Apart from the hump in the data, we can still observe a plateau in $\kappa_{x y} / T$ in the relatively higher temperature regime, which corresponds approximately with the quantized value $\pi / 12$ [see Figs.~\ref{fig:ppip-edge-overlap-invest}(a)(b)]. The size of the hump and the length of the approximately quantized plateau depends on the strength of the overlap effect. These numerical observations are in agreement with the foregoing analysis.

\subsection{Remarks}

From the preceding numerical results, one can come to the following conclusions regarding the choice of the lattice geometry: 
(i) Due to the discretization effect and bulk-state contributions, the system circumference $N_y$ must be sufficiently large, which usually has to reach a scale of $\sim 10^2$ unit cells; otherwise, a quantized plateau in $\kappa_{x y} / T$ would not appear. 
For interacting systems, calculations with such a large $N_y$ are generally impractical, unless one uses an infinite algorithm to avoid a finite-size calculation. 
(ii) Due to the potential overlap effect between edge modes, there is a constraint on the system length $N_x$. 
However, from our numerical results, one may infer that a small $N_x \leqslant 10$ is sufficient to capture the central region of the topological phase, while the precise determination of the phase boundaries remains evasive with reachable choices of $N_x$. 
This suggests that an infinite tensor-network algorithm designed for an infinite strip would be suitable for the edge current calculation.

\section{Applications}\label{sec:applications}

As preliminary applications and benchmark examples, we apply the edge current approach to two relatively challenging non-interacting systems, the Hofstadter model and the disordered $p + \mathi p$ superconductor. 
As suggested from the previous section, to mimic the lattice geometry of an infinitely long strip with a finite width, in our non-interacting calculations, we put the systems on a cylinder with a sufficiently large circumference $N_y$ and a relatively small length $N_x$. 

\subsection{Hofstadter model}

The Hofstadter model{~\cite{hofstadter1976energy}} describes the behavior of charged particles in a magnetic field, in which we intend to test the performance of the edge current approach in the case of higher Chern numbers---in that case, there are more than one edge modes at each boundary.
The Hamiltonian of the Hofstadter model is given by
\begin{equation}
  H = - \sum_{m, n} \left( f_{m, n + 1}^{\dagger} f_{m, n} + e^{\mathrm{i} 2 \pi
  \Phi n} f_{m + 1, n}^{\dagger} f_{m, n} + \mathrm{h.c.} \right) - \mu
  \sum_{m, n} f_{m, n}^{\dagger} f_{m, n},
\end{equation}
where we have used the Landau gauge. 
$\Phi = p / q$ is a rational number, which is determined by the magnetic flux through each plaquette. 
When $q > 1$, the energy band splits into $q$ subbands, each of which has a nonzero Chern number.
When the system has boundaries, there will be edge modes between the subbands, and the number of edge modes is determined by the total Chern number $C$ of the valence bands. 
Since in the Hofstadter model, the edge modes are complex (Dirac) fermions, each of these will contribute a chiral central charge $c_- = \pm 1$, where the sign depends on the chirality of the edge mode.
In our present convention, a positive chiral central charge corresponds to the counterclockwise flowing edge current, which in turn corresponds to a negative Chern number, i.e., $c_- = -C$.

In the following, by choosing different values of the magnetic flux $\Phi$ and the chemical potential $\mu$, we obtain systems with different Chern numbers.
In each choice of parameters, the total Chern number of the valence bands is calculated with the Diophantine equation{~\cite{thouless1982quantized}},
\begin{equation}
  r = q s_r + p t_r, \; | t_r | < q / 2, \; s_r, t_r \in \mathbb{Z},
  \label{eq:diophanitine}
\end{equation}
where the total Chern number of the $r$ valence bands is given by $t_r$. 
Using Eq.{~\eqref{eq:dIdT-to-kappa}}, we numerically calculate $\kappa_{xy} / T$, which is compared with the theoretical prediction given by Eqs.{~\eqref{eq:quantized_thermal_hall_conductance_expr} and \eqref{eq:diophanitine}}.
The lattice geometry, as we have stated above, is a cylinder with sufficiently large circumference $N_y=400$ and relatively small length $N_x=12, 16, 20, 24$. 

\begin{figure}[!htb]
  \resizebox{8.61cm}{!}{\includegraphics{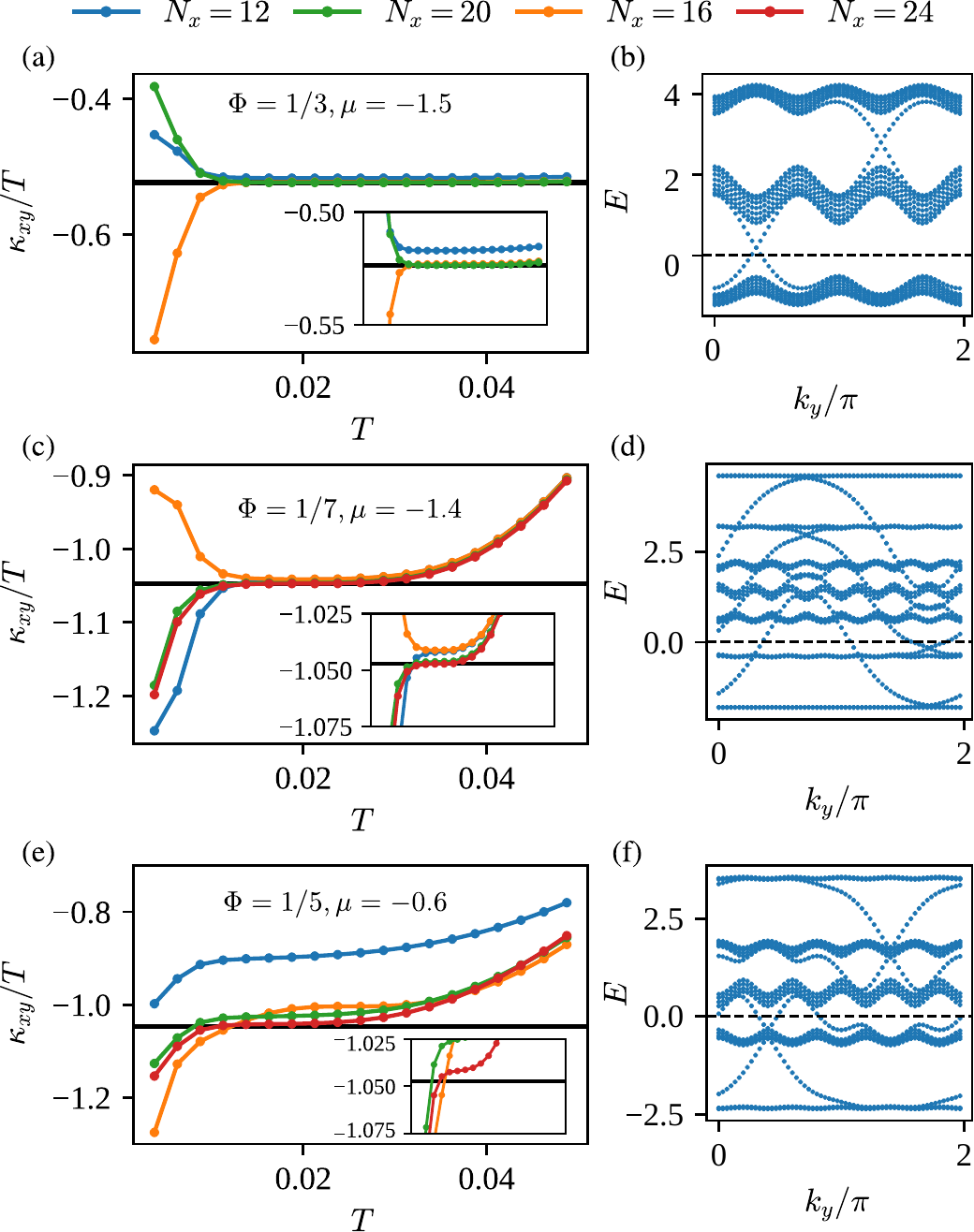}}
  \caption{Left: $\kappa_{x y} / T$ calculated in the Hofstadter model using the edge current approach for different choices of parameters and different Chern numbers, compared with the theoretically predicted quantized value $\pi c_- / 6$, which is marked as a black horizontal line. 
  The system circumference is set as $N_y = 400$, while the length of the system varies among $N_x = 12, 16, 20, 24$. 
  Right: the corresponding energy spectrum is plotted, where the system size is chosen as $N_x = 24, N_y = 60$. The numerical results of $\kappa_{x y} / T$ with (a) $\Phi = 1 / 3, \mu = - 1.5$, (c) $\Phi = 1 / 7, \mu = - 1.4$, and (e) $\Phi = 1 / 5, \mu = - 0.6$, and the energy spectrum are shown in (b)(d)(f), respectively. \label{fig:kappa-hofst}}
\end{figure}

The calculation results are shown in Fig.~\ref{fig:kappa-hofst}.
For clarification, we also correspondingly plot the energy spectrum of the system, using the same parameters $\Phi$ and $\mu$.
The energy spectrum is obtained with the lattice on a cylinder with $N_x=24$ and $N_y=60$.

First, as the simplest case, Fig.~\ref{fig:kappa-hofst}(a) shows $\kappa_{x y} / T$ calculated with $\mu = - 1.5$ and $\Phi = 1 /3$, where the Chern number equals $C = 1$, and there exists only one complex fermion mode at each boundary [see Fig.~\ref{fig:kappa-hofst}(b)]. 
In this case, we expect $c_- = - 1$. 
As shown in Fig.~\ref{fig:kappa-hofst}(a), for each of the $N_x$'s that we have considered, a plateau corresponding to the expected quantization value $ - \pi / 6$ is observed. 
The low-temperature deviations can be attributed to the discretization of edge spectra, which may be in conjunction with the overlap effect, as discussed in Sec.~\ref{sec:discussions}~B.
We also note that for $N_x = 12$, the plateau in the $\kappa_{x y} / T$ data is only approximately quantized, which indicates the presence of weak overlap between edge modes.

Secondly, we show the results in the cases with the higher Chern number $C=2$, where two electron modes flow at each boundary.
The chiral central charge becomes $c_- = - 2$, and the quantization value of $\kappa_{x y} / T$ becomes $- \pi / 3$. 
In Fig.~\ref{fig:kappa-hofst}(c), we show the results calculated with $\mu = - 1.4, \Phi = 1 / 7$. 
The corresponding energy spectrum is plotted in Fig.~\ref{fig:kappa-hofst}(d), which shows the existence of two electron edge modes. 
In the energy spectrum, one can observe that the bulk gap is smaller than the $C = 1$ case, which leads to the inference that the finite-size effect would be more severe.
Nevertheless, although only approximately quantized, the anticipated plateaus in the $\kappa_{x y} / T$ data are still observed, even for the smallest choice of the system length $N_x = 12$.

Finally, in Fig.~\ref{fig:kappa-hofst}(e), we show results with much stronger overlap effects, where we have set $\mu = - 0.6, \Phi = 1 / 5$, and the corresponding chiral central charge is $c_- = - 2$ [see Fig.~\ref{fig:kappa-hofst}(f)]. 
Due to the stronger overlap effect, in order to observe the quantized plateau in $\kappa_{x y}$, a larger system length $N_x$ is required.
As one can observe from the numerical results, only in the data of $N_x=24$ does the expected plateau appear, and the calculations for smaller system lengths all break down in this case.

\subsection{Disordered $p + \mathrm{i} p$ model}

As the second application, we apply the edge current approach in disordered systems, where translation symmetry is broken.
Since the gapless edge modes are robust due to the protection of the bulk gap, the edge current approach is expected to remain effective in the presence of relatively weak disorder.  

\begin{figure}[!htb]
  \resizebox{6cm}{!}{\includegraphics{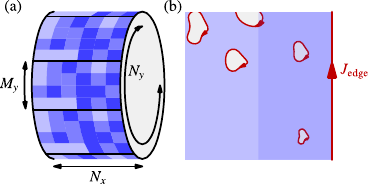}}
  \caption{(a) The setup of the supercells in the lattice system where we perform our calculation. (b) A schematic illustration of the disorder effects in the numerical calculation. The blue square represents a part of the system that is close to the boundary of the system, in which there exist small topologically trivial domains induced by the disorder. There also exist edge currents at the boundaries of these domains. The shaded region denotes where we calculate the current operator $\hat{j}^{(r)}$ defined in Eq.~(\ref{eq:j-r-def}). \label{fig:disorder-illustration}}
\end{figure}

\begin{figure}[!htb]
  \resizebox{7cm}{!}{\includegraphics{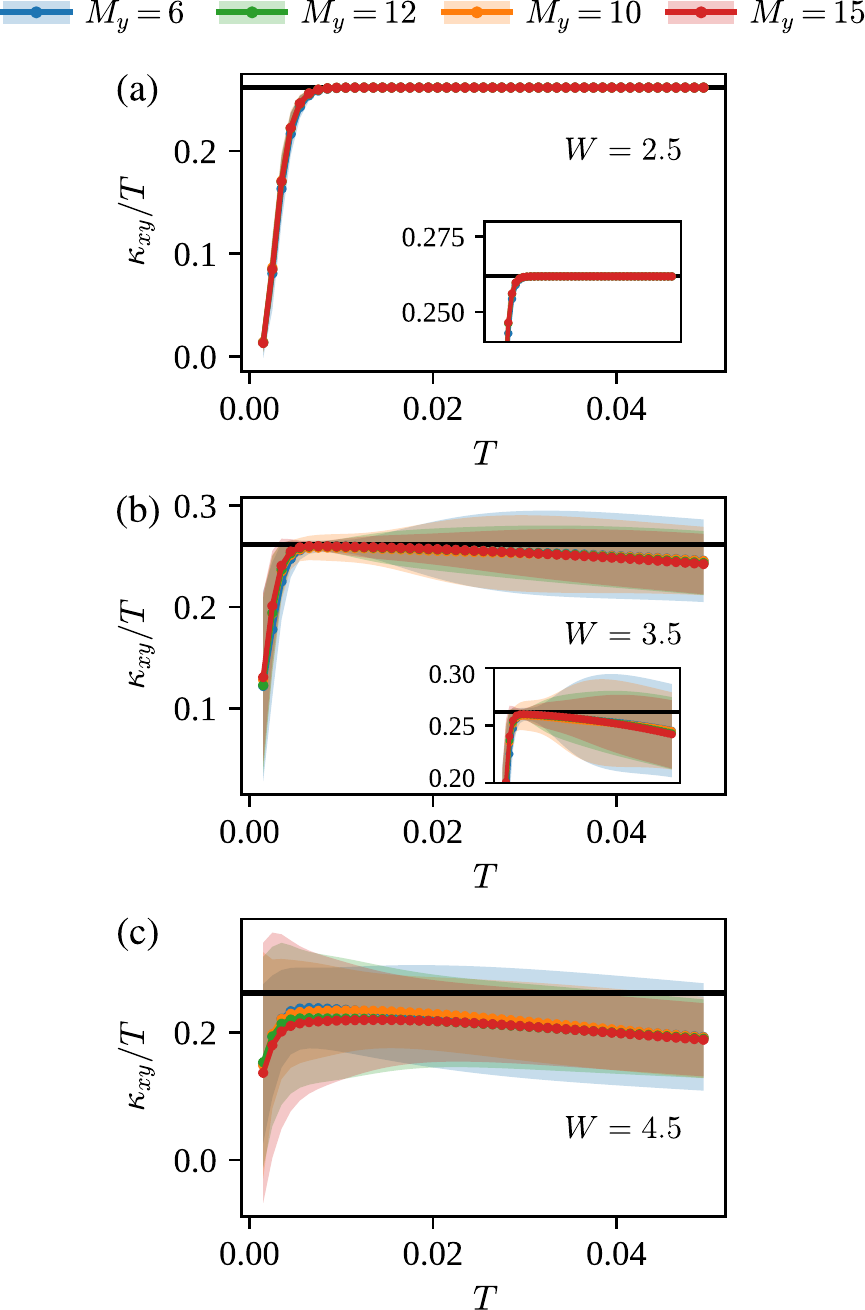}}
  \caption{$\kappa_{x y}/T$ vs $T$ plotted for different disorder strength (a) $W = 2.5$, (b) $W = 3.5$, and (c) $W = 4.0$. 
  Results calculated with different supercell sizes ($M_y=6,10,12,15$) are compared.
  The anticipated quantization value $\pi / 12$ is marked as a black horizontal line in the figure.
  The shaded region in the figure represents the error bar of the calculation results. 
  \label{fig:weak-disorder-wppip}}
\end{figure}

As a numerical test, we add on-site disorder to the clean $p + \mathrm{i} p$ model. 
The total Hamiltonian is expressed as $H_{\mathrm{tot}} = H + H'$, where $H$ is the original $p + \mathrm{i} p$ Hamiltonian Eq.~(\ref{eq:hamilt-ppip}), and $H' = - \sum_{m, n} W_{m, n} f_{m, n}^{\dagger} f_{m, n}$ represents the on-site disorder. 
Here $m, n$ are site indices, $W_{m, n}$'s are site-dependent random potentials that are uniformly distributed within the range $[-W, W]$, with $W$ representing the disorder strength. 
We fix the parameters of the original $p + \mathrm{i} p$ Hamiltonian to be $t = \Delta = 1.0$ and $\mu = 2.0$, in which case the bulk gap is $\Delta_{\mathrm{bulk}} = 2.0$. 
The lattice is put on a cylinder with circumference $N_y=600$ and length $N_x=12$. 
We gradually increase the disorder strength $W$, and for each $W$, we calculate the thermal Hall conductance $\kappa_{x y}$ by averaging over 200 different disorder configuration. 
To minimize the computational cost, we make use of a supercell trick, i.e., the system is constituted by supercells of size $N_x \times M_y$, and the disorder configuration in each supercell is identical, as schematically depicted in Fig.~\ref{fig:disorder-illustration}.
We compare results obtained with different sizes of supercells, $M_y = 6, 10, 12, 15$, and we expect the result of $\kappa_{x y}$ to converge if $M_y$ is sufficiently large.
The numerical results are shown in Fig.{~\ref{fig:weak-disorder-wppip}}. 

First, in Fig.~\ref{fig:weak-disorder-wppip}(a) we show the results calculated with disorder strength $W=2.5$, from which the anticipated quantized plateau is clearly observed. 
Furthermore, the error bars in the numerical results are almost negligible, which indicates that the disorder-induced fluctuation is very small.
Although, as one may notice, the disorder strength is slightly over the bulk gap $\Delta_{\mathrm{bulk}}=2.0$, which indicates that the disorder may induce domains with different topological properties in the system, these domains are believed to be very small and cannot lead to considerable effects on our results.
This result verifies the robustness of the edge current approach in the presence of weak disorder. 

Secondly, Fig.~\ref{fig:weak-disorder-wppip}(b) shows the result calculated with the disorder strength $W=3.5$. 
As can be observed, the disorder-induced fluctuations substantially increase and the quantized plateau is evidently diminished by the disorder. 
The reason for these fluctuations is twofold:
(i) Since the disorder strength approaches the phase transition point, the domains in the system largely increase, while the system size remains the same. 
In this regard, the overall topological property of the finite-size system may become dependent on the different disorder configurations. 
(ii) In our calculation, we need to divide the system into two parts along some central line, and some disorder-induced domains may cross the central line, which gives rise to additional edge current contributions, as schematically shown in Fig.~\ref{fig:disorder-illustration}(b).
When the sizes of the domains are sufficiently large, these additional edge currents can lead to substantial corrections to the numerical results.
Moreover, since the domain distributions vary drastically with the disorder configurations, the additional edge contributions also give rise to large fluctuations in the results, even when the overall topological property of the system persists.
These additional corrections may pose a limitation on the application of the edge current approach in the presence of strong disorder.
Nevertheless, the disorder-induced fluctuations in the plateau regime are still visibly smaller [see Fig.~\ref{fig:weak-disorder-wppip}(b)] and the plateau is still approximately quantized, which highlights the robustness of the edge current against disorder influences.

Finally, in Fig.~\ref{fig:weak-disorder-wppip}(c) with the disorder strength $W=4.0$, the plateau totally disappears and large fluctuations dominate in the whole temperature regime. 
We infer that in this case the system is quite close to phase transition and the disorder-induced fluctuations predominate. 
We also expect that in this case, the effective overall bulk gap is very close to zero, which leads to delocalization of the edge modes. 
As a result, the edge current approach is no longer applicable to this very strong disorder strength.

\begin{figure}[!htb]
  \resizebox{8.5cm}{!}{\includegraphics{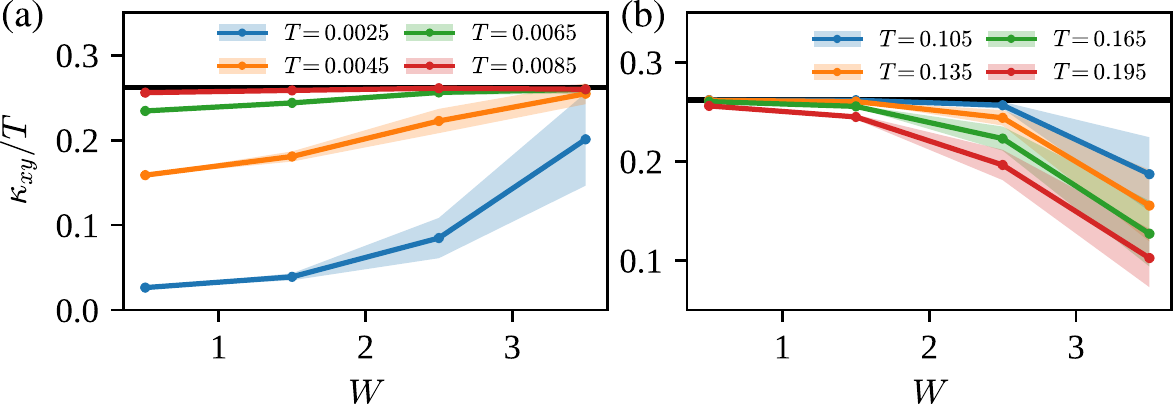}}
  \caption{$\kappa_{x y} / T$ versus the disorder strength $W$ in (a) the low-temperature regime and (b) high-temperature regime. 
  The system size is $N_x=12$ and $N_y=600$, and the supercell size is fixed as $M_y=15$, which is the largest supercell size used in our numerical calculation.
  The shaded regions represent the error bar of the numerical results, and the black horizontal line denotes the expected quantization value $\pi / 12$. 
  \label{fig:weak-disorder-wppip-vplot}}
\end{figure}

Other than the quantized thermal Hall conductance, we can also obtain some other relevant information from the numerical result of $\kappa_{x y} / T$. 
In Fig.~\ref{fig:weak-disorder-wppip-vplot}, we plot the $\kappa_{x y} / T$ with respect to the disorder strength in the low-temperature and high-temperature regime respectively.
Fig.~\ref{fig:weak-disorder-wppip-vplot}(a) shows that the deviation from the quantization at low temperatures is reduced by the weak disorder.
Recall that this deviation originates from the violation of the condition $T \gg v / N_y$, thus we infer that the effective velocity of the edge mode is reduced due to the obstruction of the disorder.
On the other hand, Fig.~\ref{fig:weak-disorder-wppip-vplot}(b) shows that $\kappa_{x y} / T$ deviates more as one increases the disorder strength in the high-temperature regime. 
This numerical observation indicates that the overall effective bulk gap is reduced by the disorder.

\section{Conclusion and outlook}\label{sec:conclusion_and_outlook}

To conclude, we have proposed an approach to calculate the quantized thermal Hall conductance in gapped chiral topological systems---calculating the thermal Hall conductance by utilizing edge thermal currents directly.
As we have seen, in contrast to the bulk Kubo formula paradigm, our edge current approach is physically transparent and hence free from subtleties such as energy magnetization corrections. 
More appealingly, the edge current approach only involves thermodynamic calculations of locally defined operators, which may get into the capability of prevailing numerical methods for interacting systems.
As a preliminary study, using several non-interacting fermion systems as platforms, we investigate the major finite-size effects that would arise in our approach, based on which we suggest that an infinite (or sufficiently long) strip would serve as the best lattice geometry in these calculations, and the capability of this approach lies at the maximal width of the strip that is reachable in numerics.
According to our numerical results, the edge current approach is competent when the system is deep inside the topological phase, but it lacks the capability to precisely determine phase boundaries.
While it still requires more progress to apply the finite-temperature tensor-network algorithms{~\cite{verstraete_matrix_2004,white_2009_minimally,stoudenmire_2010_minimally,chen_2018_exponential,kshetrimayum_2019_tensor}} to interacting systems with the suggested infinite-strip lattice geometry, our work drastically reduces the computational barrier and serves as an important step towards that goal.

As an outlook, our edge current approach can serve as a numerical instrument for the detection of chiral topological phase in microscopic models.
Specifically, a possible application lies in the thermal Hall effect in the Kitaev material $\alpha$-RuCl$_3${~\cite{kasahara_majorana_2018}}.
In addition to the recent theoretical interest in the interplay between acoustic phonons and the chiral Majorana edge mode{~\cite{vinkler-aviv_approximately_2018,ye_quantization_2018}}, one may find several other experimental features intriguing---for example, the sensitivity of the field-induced spin liquid phase upon the direction and strength of an external magnetic field{~\cite{baek_evidence_2017,zheng_gapless_2017,kasahara_majorana_2018}}.
From this perspective, it is desirable to directly compute the thermal Hall conductance in the several proposed microscopic models for $\alpha$-RuCl$_3$ and compare with the experimental observations.
Moreover, it would also be interesting to numerically examine and investigate the thermal Hall effect using the edge current approach in other theoretically proposed chiral spin liquid systems---e.g., Refs.~\cite{kalmeyer_equivalence_1987,nielsen_local_2013,bauer_chiral_2014}, as well as fractional quantum Hall systems{~\cite{kane_quantized_1997}}.

\section*{Acknowledgments}

We thank W.~Brenig, H.~Jiang, J.~R.~Shi, Q.~F.~Sun, Y.~Wan, and Y.~H.~Wu for helpful discussions. 
We acknowledge financial support from NBRPC under Grants No.~2015CB921102, No.~2017YFA0303301, and NSFC under Grants No.~11534001, No.~11504008 (W.T. and X.C.X.), Ministry of Science and Technology of China under Grant No.~2016YFA0302400 (L.W.) and the DFG through project A06 (H.H.T.) of SFB 1143 (project-id 247310070). 
W.T. acknowledges Institute of Theoretical Physics at TU Dresden for hospitality and Graduate School of PKU for financial support during part of this work.

\appendix\section{Derivation of the edge energy current of chiral topological systems}\label{app:finite-size-correction-oneoverL2}

In this appendix, we derive the edge energy current of a chiral topological system using the partition function of the edge CFT.
Along with the renowned result of Ref.{~\cite{cappelli_thermal_2002}}, which reveals the $T^2$-dependence of the edge energy current at finite temperature, we also derive the ground-state edge current, which gives rise to the finite-size correction $O(1/N_y^2)$ in Eq.{~\eqref{eq:gs_subtracted_jr}}.

We consider the chiral topological system on a cylinder, as depicted in Fig.{~\ref{fig:construction_curr_operator}}(b). 
At the two open boundaries of the cylinder there are the gapless chiral edge modes, which are described by the edge CFT.
In the edge CFT, the energy current density is defined as the thermal average of the momentum density,
\begin{equation}
  j_Q \equiv \langle \mathcal{P} \rangle_T = \frac{v^2}{2 \pi} \langle
  \mathcal{T}- \overline{\mathcal{T}} \rangle_T,
\end{equation}
where $\mathcal{T} (\omega)$ and $\overline{\mathcal{T} (\omega)}$ are the two chiral components of the energy-momentum tensor. 
Here $\omega = v \tau + \mathrm{i} x$ is a complex number, where $v$, $\tau$, and $x$ denote the velocity, the imaginary time, and the spatial coordinate, respectively.
Due to the periodic boundary condition in the spatial dimension, $\omega$ is a coordinate on a spacetime cylinder. 
Since the thermal current density is translationally invariant, it is equal to the spatial average of the momentum density,
\begin{eqnarray}
  j_Q & = & \frac{v^2}{N_y} \int_{- \mathrm{i} N_y / 2}^{\mathrm{i} N_y / 2} \frac{\mathd
  \omega}{2\pi\mathrm{i}} \langle \mathcal{T} (\omega) - \overline{\mathcal{T}
  (\omega)} \rangle_T \nonumber\\
  & = & \frac{v^2}{N_y} \langle (L_{- 1} - \overline{L}_{- 1})_{\mathrm{cyl}}
  \rangle_T, 
\end{eqnarray}
where $L_n (\overline{L}_n)$ ($n\in \mathbb{Z}$) is the Laurent mode of the energy-momentum tensor $\mathcal{T} (\omega)$($\overline{\mathcal{T} (\omega)}$), and $N_y$ denotes the circumference of the original system, which corresponds to the length of the edge theory [see Fig.{~\ref{fig:construction_curr_operator}}(a)]. 
By $z = \exp (2 \pi \omega / N_y)$, we map the spacetime cylinder to a complex plane. 
Then
\begin{equation}
  (L_{- 1})_{\mathrm{cyl}} = \frac{2\pi}{N_y} \left[ (L_0)_{\mathrm{plane}} -
  \frac{c}{24} \right] .
\end{equation}
In the following, we focus on the complex plane and the subscript ``plane'' will be dropped.

With regard to the chiral edge CFT, we only consider the simplest case, in which the anti-holomorphic part is completely absent---i.e., we assume the absence of counterflowing edge modes (the discussions below can be easily generalized to the case with counterflowing edge modes being present).
The edge CFT partition function then only contains the holomorphic part, which is given by the character $\chi_a (\tau)= \mathrm{Tr}_a (q^{L_0 - c_- / 24})$, where $c_-$ represents the chiral central charge, $q = \mathrm{exp} (2 \pi \mathrm{i} \tau)$, $\tau = \mathrm{i} v \beta / N_y$ (not to be confused with the ``$\tau$'' in $\omega = v \tau + \mathrm{i} x$), and $\beta=1/T$ represents the inversed temperature. 
Here $a$ labels the primary state of the conformal tower, which is associated with the quasiparticles in the bulk, and the trace is over the conformal tower of states~\cite{francesco2012conformal}.
By writing $\tau = v (\gamma + \mathrm{i} \beta) / N_y$, where $\gamma \rightarrow 0^+$, we then have
\begin{eqnarray}
  j_Q & = & \frac{2\pi v^2}{N_y^2} \langle L_0 - \frac{c_-}{24} \rangle_T \nonumber\\
  & = & \frac{2\pi v^2}{N_y^2} \frac{1}{\chi_a} \mathrm{Tr}_a \left( \left( L_0 -
  \frac{c_-}{24} \right) q^{L_0 - c_- / 24} \right) \nonumber\\
  & = & - \frac{\mathrm{i} v}{N_y} \left. \frac{\partial \ln \chi_a}{\partial
  \gamma} \right|_{\gamma = 0} .  \label{eq:JQexpressionCFT}
\end{eqnarray}
By introducing $\xi \equiv - 2 \pi \mathrm{i} \tau = (- 2 \pi \mathrm{i} v / N_y) (\gamma + \mathrm{i} \beta) \xrightarrow{\gamma \rightarrow 0} 2 \pi v \beta / N_y$, Eq.{~\eqref{eq:JQexpressionCFT}} can be simplified as 
\begin{equation}
 j_Q = - \frac{2\pi v^2}{N_y^2} \partial_{\xi} \ln \chi_a, \label{eq:JQexpressionCFT2}
\end{equation}
With Eq.~(\ref{eq:JQexpressionCFT2}), we can readily compute the thermal current density $j_Q$ in the two different limits $T \gg v / N_y$ and $T \rightarrow 0^+$, respectively.

When $T \gg v / N_y$, by making use of the modular transformation $\chi_a (\tau) = \sum_b S_{a b} \chi_b ( - 1/\tau)$, we get
\begin{equation}
  \ln \chi_a = \ln \left( \sum_b S_{a b} \chi_b \left( - \frac{1}{\tau}
  \right) \right) = \ln (S_{a I} \mathfrak{q}^{- c_- / 24} + O
  (\mathfrak{q}^{1 - c_- / 24})),
\end{equation}
where $\mathfrak{q} \equiv \exp (- 4 \pi^2 / \xi)$, and $S$ represents the $S$-modular matrix. 
Under the condition $N_y \gg v\beta$, $\xi = 2 \pi v \beta / N_y\rightarrow 0^+$, then $\mathfrak{q} \rightarrow 0^+$. 
Since $\mathfrak{q}$ exponentially vanishes as $\xi$ approaches zero, we infer there are no polynomial corrections in the expression of $\ln \chi_a$.
Therefore,
\begin{equation}
  \ln \chi_a = \ln (S_{a I} \mathfrak{q}^{- c_- / 24}) 
  = \frac{c_- \pi^2}{6 \xi} + \ln S_{aI},
\end{equation}
up to some exponentially decaying terms as $\xi \rightarrow 0^+$. 
According to Eq.{~\eqref{eq:JQexpressionCFT2}}, we get
\begin{equation}
  j_Q (T) = \frac{\pi c_-}{12} T^2 . \label{eq:starting-point-verified}
\end{equation}

On the other hand, we can take the zero temperature limit $T \rightarrow 0^+$, where $\xi = 2\pi v \beta / N_y \rightarrow + \infty$, and $q = \exp (- \xi)\rightarrow 0^+$ vanishes exponentially as $\xi$ approaches infinity.
In this case, the character $\chi_a$ is dominated by the contribution of the primary state $a$, i.e., $\chi_a = q^{h_a-c_-/24} + O(q^{h_a + 1 - c_- / 24})$, where $h_a$ denotes the conformal weight of the primary state. 
By making use of Eq.{~\eqref{eq:JQexpressionCFT2}}, we get
\begin{equation}
  j_Q (T = 0) = \frac{2 \pi v^2}{N_y^2} \left( h_a - \frac{c_-}{24} \right), 
\label{eq:ground-state-contribution}
\end{equation}
up to some exponentially decaying terms. 

\section{Majorana basis for non-interacting fermion systems}\label{appendix:majorana-basis}

Here, we provide the details of our calculations in non-interacting spinless fermion systems, where we make use of the Majorana basis.
The generalization to systems with multicomponent fermions at each site is straightforward.
At each lattice site $i$, we introduce two Majorana fermions $f_l^{\dagger} = \frac{1}{2} (\gamma_{l 1} + \mathrm{i} \gamma_{l 2})$ and $f_l = \frac{1}{2} (\gamma_{l 1} - \mathrm{i} \gamma_{l 2})$ for the spinless fermion mode, where $\gamma_{l, \eta}$ $(\eta = 1, 2)$ is a Majorana operator that satisfies $\gamma_a^{\dagger} = \gamma_a$ and $\{ \gamma_a, \gamma_b \} = 2 \delta_{a b}$. 
Generally, under this basis, one can represent any quadratic Hamiltonian with $N$ sites using a $2 N \times 2 N$ real antisymmetric matrix, which is suitable to describe non-interacting fermion systems.

\subsection{Thermodynamic calculation under the Majorana basis}

For the quadratic Hamiltonian $\hat{H} = (\mathrm{i} / 4) \sum_{a, b} \gamma_a M_{a b} \gamma_b$, where $M$ is a $2 N \times 2 N$ real antisymmetric matrix, $\gamma_a$ is the Majorana fermion operator, and $a, b$ are the indices of Majorana fermions. Any $2 N \times 2 N$ antisymmetric matrix $M$ can be block-diagonalized, with an orthogonal transformation, into the following form:
\begin{equation}
  \tilde{M} = O M O^T = \bigoplus_{l = 1}^N \left( \begin{array}{cc}
    0 & \lambda_l\\
    - \lambda_l & 0
  \end{array} \right),
\end{equation}
where $O$ is the orthogonal matrix, and $\lambda_l > 0$. By introducing the new Majorana fermion basis $\tilde{\gamma} \equiv O \gamma$, the Hamiltonian is expressed as $\hat{H} = (\mathrm{i} / 4) \sum_{a, b} \tilde{\gamma}_a \tilde{M}_{a b} \tilde{\gamma}_b$, and we have
\begin{equation}
  \exp (- \beta \hat{H}) = \exp \left( - \frac{\mathrm{i} \beta}{4}
  \widetilde{\gamma}^T \tilde{M} \widetilde{\gamma}
  \right) = \prod_{l = 1}^N \exp \left( - \frac{\mathrm{i}\beta}{2} \lambda_l
  \tilde{\gamma}_{l, 1} \tilde{\gamma}_{l, 2} \right),
\end{equation}
where we have introduced $\gamma = (\gamma_{1, 1}, \gamma_{1, 2}, \gamma_{2, 1}, \ldots, \gamma_{N, 2})^T$. By introducing the complex fermion $a_l = (\tilde{\gamma}_{l, 1} - \mathrm{i} \tilde{\gamma}_{l, 2})/2$ and $a_l^{\dagger} = (\tilde{\gamma}_{2 l - 1} + \mathrm{i} \tilde{\gamma}_{2 l})/2$, we have $\mathi \tilde{\gamma}_{2 l - 1} \tilde{\gamma}_{2 l} = - 2 a^{\dagger}_l a_l + 1$ and then 
\begin{equation}
  \exp (- \beta \hat{H}) = \prod_{l = 1}^N \exp \left[ \beta \lambda_l \left(
  a_l^{\dagger} a_l - \frac{1}{2} \right) \right] \dot{=} \bigotimes_{l = 1}^N
  \left( \begin{array}{cc}
    \mathrm{e}^{- \frac{1}{2} \beta \lambda_l} & \\
    & \mathrm{e}^{\frac{1}{2} \beta \lambda_l}
  \end{array} \right), \label{eq:majorana-expminusbetaH}
\end{equation}
from which one can calculate the partition function as
\begin{equation}
  Z = \mathrm{Tr} (\mathrm{e}^{- \beta \hat{H}}) = \prod_{l = 1}^N 2 \cosh \left(
  \frac{1}{2} \beta \lambda_l \right) .
\end{equation}

Under the Majorana basis, for our purposes, it is useful to introduce the correlation matrix $K$ as
\[ K_{a b} = \frac{\mathrm{i}}{2 Z} \mathrm{Tr} (\mathrm{e}^{- \beta \hat{H}}
   [\gamma_a, \gamma_b]) . \]
Under the orthogonal transformation $O$, the correlation matrix transforms as
\begin{equation}
  K = O^T \tilde{K} O,
\end{equation}
where $\tilde{K}_{d e} = (\mathrm{i} / 2 Z) \mathrm{Tr} (\mathrm{e}^{- \beta \hat{H}} [\tilde{\gamma}_d, \tilde{\gamma}_e])$. One can easily verify that $\tilde{K}$ is a $2 \times 2$-block diagonalized matrix, with nonvanishing elements given by
\begin{eqnarray}
  \tilde{K}_{(l, 1), (l, 2)} & = & \frac{\mathrm{i}}{Z} \mathrm{Tr} (\mathrm{e}^{-
  \beta \hat{H}} \tilde{\gamma}_{l, 1} \tilde{\gamma}_{l, 2}) \nonumber\\
  & = & - \frac{1}{Z} \mathrm{Tr} [\mathrm{e}^{- \beta \hat{H}} (2 a_l^{\dagger}
  a_l - 1)] \nonumber\\
  & = & - \mathrm{{\frac{1}{Z}}Tr} \left[ \bigotimes_{m = 1}^N \left(
  \begin{array}{cc}
    \mathrm{e}^{- \frac{1}{2} \beta \lambda_m} & \\
    & \mathrm{e}^{\frac{1}{2} \beta \lambda_l}
  \end{array} \right) \left( \begin{array}{cc}
    1 - 2 \delta_{l m} & \\
    & 1
  \end{array} \right) \right] \nonumber\\
  & = & - \tanh \left( \frac{1}{2} \beta \lambda_l \right), 
\end{eqnarray}
and $\tilde{K}_{(l, 2), (l, 1)} = - \tilde{K}_{(l, 1), (l, 2)}$, where $l = 1, 2, \ldots, N$. In the zero temperature limit, $\tilde{K}_{2 k - 1, 2 k} = - \tilde{K}_{2 k, 2 k - 1} = - 1$.

With the correlation matrix $K$, one is able to calculate the thermal average of an observable $\hat{Q} = (\mathrm{i} / 4) \sum_{a, b} \gamma_a M^{(Q)}_{a b} \gamma_b$ as
\begin{equation}
  \langle \hat{Q} \rangle = \frac{\mathrm{i}}{4 Z} \sum_{a, b} M^{(Q)}_{a b}
  \mathrm{Tr} (\mathrm{e}^{- \beta \hat{H}} c_a c_b) = - \frac{1}{4} \mathrm{Tr}
  (M^{(Q)} K),
\end{equation}
where $M^{(Q)}$ is the real antisymmetric matrix corresponding to the quadratic Hermitian operator $\hat{Q}$.

\subsection{Energy current operator in the Majorana basis}

Generally speaking, the energy current operator defined in Sec.~\ref{sec:method_sec} is of the form of commutators of local Hamiltonian terms, i.e.,
\begin{equation}
  \hat{J}_{k \rightarrow j} = - \mathrm{i} [\hat{H}_j, \hat{H}_k],
\end{equation}
where $j, k$ are indices of local Hamiltonians. Under the Majorana basis, we introduce $\hat{H}_j = (\mathrm{i} / 4) \sum_{a, b} \gamma_a M_{a b}^{(j)} \gamma_b$, where $M^{(j)}$ is the antisymmetric Hamiltonian corresponding to the local Hamiltonian $\hat{H}_j$. By a straightforward calculation, we get
\begin{eqnarray*}
  {}[\hat{H}_j, \hat{H}_k] & = & - \frac{1}{16} \sum_{a, b, d, e} M_{a
  b}^{(j)} M_{d e}^{(k)} [\gamma_a \gamma_b, \gamma_d \gamma_e]\\
  & = & - \frac{1}{8} \sum_{a, b, d, e} M_{a b}^{(j)} M_{d e}^{(k)}
  (\delta_{b d} \gamma_a \gamma_e - \delta_{b e} \gamma_a \gamma_d
  \\
  &  &  + \delta_{a d} \gamma_e \gamma_b - \delta_{a e} \gamma_d
  \gamma_b),
\end{eqnarray*}
and after rearrangement of terms, we find
\begin{equation}
  [\hat{H}_j, \hat{H}_k] = - \frac{1}{4} \sum_{a, b} \gamma_a [M^{(j)},
  M^{(k)}]_{a b} \gamma_b .
\end{equation}
Therefore, we get the antisymmetric matrix representation corresponding to the energy current operator $\hat{J}_{k \rightarrow j}$,
\begin{equation}
  M^{(J, k \rightarrow j)} = - [M^{(k)}, M^{(j)}],
\end{equation}
and the thermal average of $\hat{J}_{k \rightarrow j}$,
\begin{equation}
  \langle \hat{J}_{k \rightarrow j} \rangle = \frac{1}{4} \mathrm{Tr}
  ([M^{(k)}, M^{(j)}] K),
\end{equation}
where $K$ is the correlation matrix.


\bibliography{thermalhall}
\bibliographystyle{apsrev4-1}

\end{document}